\documentclass[preprint,amsmath,prb]{revtex4}
\usepackage[dvips]{graphicx}
\usepackage[T1]{fontenc}
\newcommand{\slat}{$\textrm{(MnTe)}_{\textrm{m}} \textrm{(ZnTe)}_{\textrm{n}}$ }

\newcommand{\veck}{\vec{k}}
\newcommand{\genslat}{Mn-VI/II-VI }
\newcommand{\Ab}{$\bar{\textrm{A}}$}
\newcommand{\Bb}{$\bar{\textrm{B}}$}
\begin{document}
\title{Interlayer Exchange Coupling in \slat Superlattices}
\author{R. W\k{a}sowicz}
\affiliation{Institute of Theoretical Physics, Warsaw University, ul. Ho\.za 69, 00-681 Warszawa, Poland}
\begin{abstract}
The interlayer exchange coupling mediated by the valence band electrons in \slat superlattices has been examined theoretically. The calculations have been performed within the framework of three dimensional tight binding model. Both types of magnetic configurations in the superlattice, resulting from zinc-blende MnTe real magnetic structure, have been considered. The total energy calculations have revealed much faster decrease of the coupling strength with nonmagnetic spacer thickness than that observed in NaCl structures. In addition a new feature -- nonmonotonic character of that decrease -- has been found.
\end{abstract}

\maketitle

\section{Introduction}

Superlattice \slat belongs to the class of widely examined \genslat systems. Its zinc-blende form was obtained in the process of MBE growth in late 80-ies. Since then most of the experimental work has been focused on the magnetic structure while the theoretical calculations have tried to explain the mechanism of Mn-Mn coupling. Although the vast majority of these semiconducting superlattices turned out to be antiferromagnetic, which seriously limited the scope of possible applications, they offered the opportunity to study magnetic phenomena in the systems without free carriers. Of all the semiconducting superlattices the most interesting ones became those incorporating $\textrm{Mn}^{2+}$ and $\textrm{Eu}^{2+}$ ions. That is because the magnetic moment of the ions does not contain contribution from the orbital degrees of freedom and hence is insensitive to the crystal field. As a result, the effective Hamiltonian describing the interactions between the spins takes simple Heisenberg form (the other components possible in this structure are negligibly small). 

The experiments performed on various \genslat systems revealed rather complicated magnetic structures. Apart from the considered here MnTe/ZnTe superlattices among other examined systems were: MnTe/CdTe,\cite{NGFBSF1995} MnSe/ZnSe \cite{SKLGFRLO1991} and MnSe/ZnTe.\cite{GSLKFR1992} The analysis of the neutron-diffraction data showed two kinds of the spin ordering: the antiferromagnetic one of type III (AFM III) and the helical one originating from the former. The helical ordering was observed in MnTe/CdTe and MnSe/ZnTe superlattices. In both cases these were incommensurate phases. Giebultowicz \textit{et al.}\cite{GKSLFR1993} drew an attention to the role of strain in determining the magnetic structure. They pointed out the essential differences between the systems with built-in compressive and tensile strain. The mentioned before superlattices with helical ordering are examples of the systems with a shorter lattice parameter in pure MnTe (MnSe) than in CdTe (ZnTe). In the case of MnTe/ZnTe superlattices, studies of the thin films and semibulk samples revealed that influence of strain is manifested in different relative population of the domain types.

Further experiments focused on observations of the spin correlations between different magnetic layers in the superlattice. Such correlations were detected in both, helical \cite{NGFBSF1995} and colinear,\cite{LRFG1998} systems. Although the range of the coupling turned out to be shorter than that found in metallic superlattices, it was sufficiently long to raise the question concerning mechanism of the correlation. As free carriers are not present in the considered superlattices, models tailored for the metallic systems are inapplicable in this case. This situation has stimulated several researchers to work out the new models \cite{BK2001,R1998,SSO1998} appropriate for the semiconducting materials. The model of Blinowski and Kacman proposed in Ref.\ \onlinecite{BK2001} takes into account both band and magnetic structures of the superlattice. It consists of calculating the band structure in the whole Brillouin zone (BZ) for the two magnetic patterns and comparing the total electronic energies of the both systems. According to the authors, this scheme, in spite of neglecting many-body effects during the calculation of the total energy, should lead to the correct results, as only small spin-dependent changes in the total energy determine the configuration in the ground state. This model was succesfully applied to EuS/PbS and EuTe/PbTe systems. 
On the other hand, the model by Rusin (Ref.\ \onlinecite{R1998}) explains the interlayer exchange coupling (IEC) as a phenomenon driven by the shallow donor impurities residing inside the nonmagnetic layers.
The whole calculation is performed within the framework of the effective mass approximation with the perturbation Hamiltonian of Kondo form. The application of this scheme to MnTe/CdTe superlattices yielded results which were in accordance with the experimental observations. Although both models turned out to be succesful, it is risky to draw any decisive conclusions, as differences between various systems are significant. 

In this work we have decided to use the model of Blinowski and Kacman. One of the reasons was the lack of data concerning the presence of the shallow donors in the real MnTe/ZnTe superlattices in which the effect of IEC was observed. Although among the examined systems were MnTe/ZnTe ones doped intentionally with Cl atoms \cite{SRKLBF2000} and the enhancement of the correlation range was detected, the authors raised several questions. Firstly the electronic levels introduced by Cl atoms seemed to be the deep ones, secondly the same effect of the IEC elongation was observed in the samples with doping in the magnetic layers only. To stress it again -- as a presence of the shallow donors in undoped samples is unknown, the usage of the model which does not require any impurity to produce the coupling seemed more appropriate.

This paper is organized as follows. In section~\ref{descr} we describe both the constituents and the superlattice itself with its different magnetic structures. The next section is devoted to present the model used to perform the calculations. Section~\ref{res} includes presentation and discussion of the obtained results.

\section{Description of the system}\label{descr}

The considered superlattices consist of a sequence of alternating blocks -- nonmagnetic and magnetic ones. Both components crystallize in the zinc-blende structure, however, it should be stressed that in bulk the latter naturally forms NiAs structure. ZnTe is a well known semiconductor with a direct energy gap ($E_g=2.56$ eV) at the $\Gamma$ point. The top of its valence band consists mostly of the anion p-states while the conduction band bottom -- of the cation s-states.
 
MnTe in the zinc-blende form is much less known, mainly due to its different native structure. Moreover, as a consequence of a non-zero magnetic moment carried by each Mn cation ($\textrm{Mn}^{2+}$) one has to take into account the spin degrees of freedom to describe the structure of the system. The magnetic order in the MnTe bulk crystals was first studied by Wei and Zunger \cite{WZ1987} -- their cohesive energy first-principles calculations have confirmed that at equilibrium MnTe assumes the antiferromagnetic (AFM) NiAs structure and pointed at the AFM zinc-blende form as a more stable than the ferromagnetic (FM) one, however, these calculations considered only the simpler AFM type I form. Wei and Zunger's later calculations,\cite{WZ1993} which accounted for AFM type III structure as well found it to be a ground state of the zinc-blende MnTe. The calculations performed in Ref.\ \onlinecite{WZ1987} and Ref.\ \onlinecite{LHEC1988} indicate that FM zinc-blende MnTe should be a direct gap semiconductor with relatively narrow d-bands. The repulsion between the tellurium p and Mn d-states results in a negative value of the p-d exchange. This pushes up the majority spin (up) valence band with respect to the up d-band, thus reducing the energy gap for the majority spin. In contrast for the minority spin, the d-bands located in the energy gap repel the valence band, extending the energy gap. The calculations of the electronic band structure of the type I AFM zinc-blende MnTe indicate that this system is a semiconductor with a direct energy gap which is a bit wider than that of FM zinc-blende MnTe.\cite{WZ1987} What is even more important the AFM ordering decreases the exchange splitting of Mn 3d-states and, therefore, leads to an increase of the p-d hybridization.
 
\begin{figure}
\includegraphics[scale=0.3]{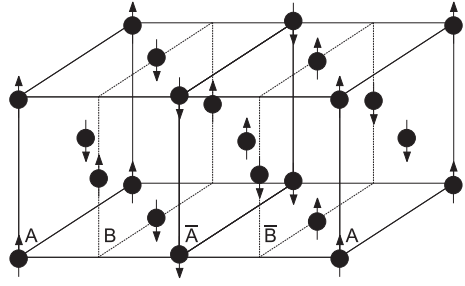}
\caption{\label{fig3} AFM Type III unit cell.}
\end{figure}

The first films of the zinc-blende MnTe, as well as various superlattices containing this magnetic semiconductor (eg. ZnTe/MnTe, CdTe/MnTe) have been obtained by molecular beam epitaxy. Since then a lot of experimental and theoretical effort has been put to explain the properties of this compound. First of all, the neutron-diffraction experiments, which enabled to study the MnTe magnetic structure, have revealed \cite{GKSLFR1993} that in the zinc-blende MnTe below N\'eel temperature all the spins are antiferromagnetically arranged within (100)-type planes and the consecutive sheets are stacked according to the scheme AB\Ab \Bb $\,$ (see Fig.\ \ref{fig3}, where \Ab $\,$ stands for `A with all the spins reversed'). This kind of order is known as type III antiferromagnetism and indicates that the nearest and next nearest neighbour exchange constants, $J_{\textrm{NN}}$ and $J_{\textrm{NNN}}$, are both antiferromagnetic. Magnetic unit cell of such a system is presented in Fig.\ \ref{fig3}. On the other hand, Larson \textit{et al.} \cite{LHEC1988} have shown theoretically that the dominant mechanism which couples the spins of the Mn-Mn pair is the antiferromagnetic superexchange (approx. 95\% contribution). Indirectly this finding draws attention to the role of the p-d hybridization in setting the MnTe magnetic structure.

Neutron-diffraction experiments have been applied to investigate not only the magnetic structure of pure MnTe, but also to examine possible correlation effects in the superlattices.\cite{LRFG1998} The authors of Ref.\ \onlinecite{LRFG1998} have measured magnetic scattering in several \slat superlattices with different, thick (n=18) and relatively thin (n=3--7), nonmagnetic spacers. As a result, the presence of the interlayer exchange coupling in the systems with 3 to 5 ZnTe monolayers was observed.

In fact two types of correlations have been found -- the so called `in-phase' and `antiphase' (or `out-of-phase') coupling. They resemble the well known ferromagnetic and antiferromagnetic couplings which may form in the system with ferromagnetic order within atomic planes. If the superlattice growth can be perceived as a process of substitution of the Mn planes in bulk MnTe by the Zn ones, then the in-phase configuration is obtained when magnetic structure is retained while the antiphase when the spin directions in every second magnetic layer are reversed. One of the ways to describe superlattice AFM III magnetic structure is to give sequence of letters A, B, \Ab, \Bb $\,$ and Z (where the first 4 letters describe inequivalent Mn planes and the last one -- Zn plane). In this language the system $\textrm{(MnTe)}_2\textrm{(ZnTe)}_3$ has the following configurations:
\begin{itemize}
\item in-phase -- ABZZZB\Ab ZZZ\Ab \Bb ZZZ\Bb AZZZ
\item antiphase -- ABZZZ\Bb AZZZ\Ab \Bb ZZZB\Ab ZZZ
\end{itemize}
However, there are essential differences between systems with even and odd number of cation planes per chemical unit cell (i.e.\ n+m). To illustrate this we give configurations for the system $\textrm{(MnTe)}_2\textrm{(ZnTe)}_4$:
\begin{itemize}
\item in-phase -- ABZZZZ\Ab \Bb ZZZZ
\item antiphase -- ABZZZZABZZZZ
\end{itemize}
As one can see, for n+m being an even number the magnetic unit cell is only doubled with respect to the chemical unit cell, while in the other case it is four times larger. Moreover, there is another important feature which distinguishes various systems, namely, the configuration of the magnetic planes bordering the nonmagnetic layer. In this case the quantity, which takes control over the configuration, is the number of nonmagnetic planes. For odd n the bordering magnetic planes are of the same kind, i.e.\ A-A, B-B, A-\Ab, B-\Bb, \Ab -\Ab $\,$ or \Bb -\Bb. It means that on both planes ions occupy equivalent positions, i.e.\ these sites can be matched by translation along growth direction only. In the other case the configurations A-B, A-\Bb $\,$ or B-\Ab $\,$ arise. The influence of these differences on the coupling strength will be discussed later. In the above considerations we have omitted anion planes which, although play crucial role in the coupling of local spins, do not contribute to the description of the magnetic structure. These planes are, obviously, located halfway between the nearest cation planes. 
 
\section{Method of calculation}

To investigate the interlayer exchange coupling in \slat system we calculate electronic structure of the given superlattice using tight-binding method. In this method one has to choose a set of orbitals for each element constituting material in question. It is well known \cite{VHD1983} that bonds of tetrahedrally coordinated semiconductors are formed by $\textrm{sp}^3$ orbitals, hence in order to reproduce bands properly these four orbitals have been taken into account for both Zn and Te ions. The situation for Mn is different. The presence of fairly well localized d orbitals, whose electrons not only build spin S=$\frac{5}{2}$ but also screen inner shells, results in less important role of p-orbitals. Instead, the orbitals of d-symmetry have to be included, as their hybridization with anion p-orbitals plays crucial role in the coupling of the localized spins. In the following we take explicitly only s orbital for Mn ion, whereas the d-orbitals have been included within the framework of second order perturbation theory. 

Having determined the content of the basis, as a next step we had to choose the range of interionic interactions. 
As we wanted to keep small number of parameters, however yielding satisfactory shape of electronic bands, we have let for next-nearest neighbour (NNN) interactions utmost but not all of them have been included.
The reason for that was awareness that it is the valence band which plays much more important role and hence the effort should be focused to improve its shape. The outcome of the described procedure is the following: all the nearest neighbour (NN) interactions have been taken into account in both constituents of the superlattice while NNN interactions have been included among the p-states of both the nonmagnetic cations and of all Te anions. Apart from the parameters describing the ion-ion interactions there are the on-site energies (e.g. $\epsilon^{(\textrm{Zn})}_s$, $\epsilon^{(\textrm{Zn})}_p$). In the case of Mn ions, the spin splitting of the s level, which results from the Coulomb exchange, has to be taken into account. This splitting $\Delta$ has not been treated as an additional parameter but assumed to be equal $0.5$ eV, a value estimated from the value of the direct exchange constant $\alpha$ for FM MnTe.

In the final step all the free parameters have been determined by fitting the eigenvalues of the obtained matrices in the high symmetry points of Brillouin zone (BZ) to the values of known band structures of ZnTe \cite{VHD1983} and FM MnTe.\cite{LHEC1988}
The parameters describing anion-anion interaction (i.e.\ NNN parameters) are present in the both sets of the parameters hence they cannot be independent, as we want to use the same parameters in the both layers of the superlattice. They have been obtained by the fit to the band structure of MnTe only.

Finally, the Hamiltonian matrix for the superlattice has been constructed based on choosen set of orbitals for each element and determined parameters. As there is some mismatch (approx. 4\%) between lattice constants for both constituents, the lattice constant for the superlattice has been assigned a mean value ($\textrm{a}_{\textrm{SL}}=0.5(\textrm{a}_{\textrm{ZnTe}}+\textrm{a}_{\textrm{MnTe}})$). On-site energies on Te anions bordering Mn and Zn cations have been treated in the same way (i.e.\ $\epsilon_{\alpha}^{\textrm{SL}}=0.5 (\epsilon_{\alpha}^{\textrm{(ZnTe)}}+\epsilon_{\alpha}^{\textrm{(MnTe)}})$). All the parameters describing ion-ion interactions have been scaled according to Harrison's formulae.\cite{H1980}

The inclusion of p-d hybridization leads to the corrections to the matrix elements between anion p-states according to the second order perturbation theory formula:
\begin{widetext}
\begin{equation}~\label{eq1}
H_{p^{\prime} j^{\prime} \sigma,p j \sigma}(\veck) \longrightarrow H_{p^{\prime} j^{\prime} \sigma,p j \sigma}(\veck)+\frac{1}{\epsilon_p-\epsilon_d} \sum_{\beta \in {\cal A},i} \langle p^{\prime} j^{\prime} \sigma, \veck|H|\beta i \sigma, \veck \rangle \langle \beta i \sigma, \veck|H|p j \sigma, \veck\rangle,
\end{equation}
\end{widetext}
where by ${\cal A}$ we denoted the five d-orbitals: $d_{xy}, d_{xz}, d_{yz}, d_{x^2-y^2}, d_{3z^2-r^2}$; $j, j^{\prime}$ mark position of Te sites, while $i$ runs over all Mn ions in the unit cell. In practice, the sum over Mn ions in the above formula is restricted only to those, which are the nearest neighbours to both $j$ and $j^{\prime}$. Energetic fraction written before the sum in Eq.\ \ref{eq1} indicates that we have neglected crystal field splitting of d-states.

As we have mentioned in the previous section, the size of the superlattice unit cell depends on the parity of n+m. For even n+m there are 4(n+m), while in the other case as many as 8(n+m), inequivalent ions in the unit cell, i.e.\, for even and odd superlattices we have to consider 20m+32n and 40m+64n orbitals, respectively. After Blinowski and Kacman\cite{BK2001} we have adopted the difference between the total energies of the in-phase and antiphase configurations as a measure of strength of the interlayer exchange coupling. The analysis of the symmetries of the Hamiltonian matrix has let reduce tetragonal Brillouin zone to its eighth part. The integration over occupied bands has been performed with the use of Simpson procedure on the 3D grid of points. In the described calculations we have neglected spin-orbit coupling (as the available data on the band structures of the zinc-blende MnTe dropped this term as well).

\section{Results}\label{res}

\begin{widetext}
\begin{figure}
\includegraphics[scale=0.7]{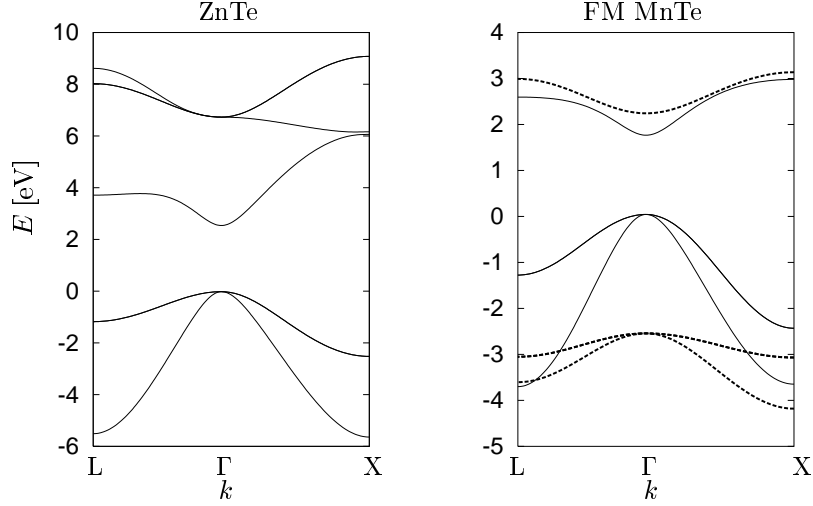}
\caption{\label{fig1}Calculated band structures of ZnTe and ferromagnetic MnTe. Dashed lines on the latter graph depict bands for minority spin.}
\end{figure}
\end{widetext}

The application of the methods described in the previous section has yielded 7 parameters for MnTe, 11 for ZnTe and the 2 common ones. The band structures obtained with these parameters are presented in Fig.\ \ref{fig1}. On both graphs the lowest lying narrow band (approx. from $-13$ eV to $-12$ eV) has been omitted for clarity reasons. In the case of ZnTe the comparison with the results of Vogl \textit{et al.} \cite{VHD1983} shows differences which do not exceed 5\% for the valence bands. Not surprisingly the situation with the conduction bands is not as good. Although the band gap, as well as the energy at $X_1$ point, are well reproduced, there is an almost 15\% difference at the $L_1$ point. Higher conduction bands are even worse matched -- the sequence of bands in $L_1$ and $L_3$ symmetry points is reversed. For MnTe the fit is not as good as in the previous case. At least two reasons may be responsible for that: (1) we had at hand only energies at $\Gamma$ and $X$ points, whereas ZnTe bands were known along $\Lambda$ line as well; (2) the set of the orbitals for Mn ion did not include the p-states, which although less important, could refine the band structure. The valence band for majority spin seems to be wider (the differences at the points $X_{3v}$ and $X_{5v}$ are around 11\% and 6\%, respectively). Smaller differences (less than 8\%) occur for the minority spin valence band, but in turn this band is narrower. The spin splitting of the valence band edge agrees well with the results of Larson \textit{et al.} \cite{LHEC1988} while that for the conduction band differs by almost 70\%. Again, this difference is a consequence of the worse treatment of the conduction bands within the framework of tight-binding model. To summarize this part -- we have obtained model band structures with main features, i.e.\ the shape and the width of the valence band, satisfactorily reproduced. 

\begin{figure}
\includegraphics[scale=0.9]{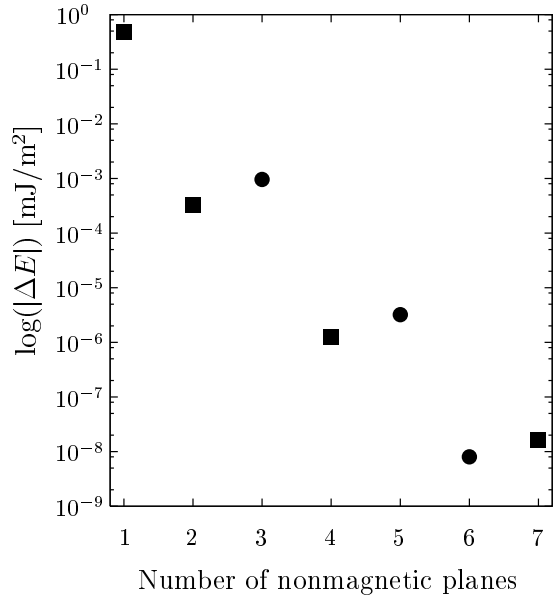}
\caption{\label{fig2} The absolute value of the energy difference between the in-phase and antiphase configurations per unit area as a function of the nonmagnetic layer thickness (squares represent these systems for which the in-phase configuration has lower energy).}
\end{figure}

We have examined a series of \slat systems with varying values of n and m. Calculations have been performed for the superlattices of both n+m parity. All the systems are semiconducting in character with the band gap around 2 eV, however, we have not verified whether this gap is direct. To calculate the total energy of the valence electrons, integrations on the grid of 1331 points in the reduced Brillouin zone have been performed. Such a choice of the points number has been determined after comparing the results obtained for different grids. 
Calculated values of the energy differences between the in-phase and antiphase configurations show that there is a slight dependence on the thickness of the magnetic layer (i.e.\ on the number m), thus superlattices \slat can be classified by the number n only. The only exceptions are superlattices with 4 and 6 nonmagnetic planes in the layer. In both cases results obtained for the systems with 2 magnetic planes differ from those calculated for the superlattices $\textrm{(MnTe)}_{\textrm{m}} \textrm{(ZnTe)}_{4}$ and $\textrm{(MnTe)}_{\textrm{m}} \textrm{(ZnTe)}_{6}$ with m=3,4,5,6. The total energy differences for these two superlattices with m=2 have the same sign but are smaller in magnitude by approx. 37\% and 20\% for the systems with n=4 and n=6, respectively, as compared to the results for m>2.
Results are presented in Fig.\ \ref{fig2} (for n=4,6 data calculated with m>2 is presented). 

Unfortunately the neutron-diffraction allows to prove the existence of the correlation but not to determine the strength of the coupling -- actually we do not know how to measure the interlayer coupling strength in the system with inplane antiferromagnetic order.
According to the calculations performed by Blinowski and Kacman \cite{BK2001} for the superlattices IV-VI the interlayer exchange coupling should vanish monotonically with the nonmagnetic layer width.
Although the calculations presented in Ref.\ \onlinecite{BK2001} concerned systems crystallizing in NaCl structure (EuS/PbS and EuTe/PbTe), for both the ferromagnetic and antiferromagnetic (AFM type II) layers almost perfect exponential falling was predected. In contrast, in our work we have obtained clear rise of the coupling strength when passing from n=2 to n=3, from n=4 to n=5 and from n=6 to n=7. It seems that this effect may be, at least partially, understood on the grounds of geometrical analysis. As we have mentioned in Sec.\ \ref{descr} there is a difference between the systems with even and odd number of nonmagnetic planes. In the systems with odd n, the nearest magnetic planes neighbouring through the nonmagnetic layer are of the same type and hence each magnetic cation can `see' exactly one nearest cation on the other side of the spacer. Contrary, if the superlattice with the bordering magnetic planes of the mixed type, i.e.\ A-B, is concerned, then every localized spin has got as many as four spins -- nearest neighbours -- across the nonmagnetic layer. Of these four spins two spins are parallel and two are antiparallel. The presence of equivalent magnetic planes, i.e.\ A-A, B-\Bb $\,$ etc.\, in the system should lead to the greater differences of the total energies between the in-phase and antiphase configurations because for one of these configurations all the pairs of the planes are coupled ferromagnetically while for the other -- antiferromagnetically. On the other hand there is not such a clear difference between the in-phase and antiphase configurations in the superlattices with an even number of nonmagnetic planes in the layer.
In spite of that one can try to recover the notions of ferromagneticlike and antiferromagneticlike couplings in this case as well. It is enough to notice that instead of considering the spin configurations on the nearest magnetic planes bordering across the nonmagnetic layer one has to take pairs of such planes. As a consequence configuration of type AB-AB can be thought of as ferromagnetic (FM) coupling, while the other, i.e.\ AB-\Ab \Bb, as antiferromagnetic (AFM) one. What is more important, for the given configuration (in-phase or antiphase) all the couplings in the superlattice are of the same type (i.e.\ either FM or AFM) just like in the case of `odd' superlattices. These observations in connection with extremely weak IEC in the superlattices with 4 and 6 nonmagnetic planes in the layer may suggest that 2 magnetic planes are not enough to stabilize the IEC in these systems. 
 
Another feature which distinguishes system under consideration from mentioned before AFM type II IV-VI superlattices is a pace of interlayer exchange coupling decrease. We can see in Fig.\ 6 in Ref.\ \onlinecite{BK2001} that IEC drops 3 orders of magnitude when spacer thickness is increased from 2 to 10 monolayers. In our work we have found that it was enough to increase the number of nonmagnetic planes from 1 to just 6 to end with almost 8 orders of magnitude drop in IEC. The obtained results and described differences between the systems with even and odd number of nonmagnetic planes in the layer, raise the question if it is not more appropriate to consider these systems separately. If so, then it can be said that in both cases exponential falling of IEC is observed. In this view types of couplings between the magnetic planes across the nonmagnetic layer in the ground state are the following:
\begin{itemize}
\item odd (1-3-5-7) -- AFM-AFM-FM-FM
\item even (2-4-6) -- FM-AFM-AFM
\end{itemize} 
In both cases type of the preferred configuration changes with the width of the nonmagnetic layer, however, these changes go in opposite directions. When analyzing these behaviours one must remember that the abbreviations `FM' and `AFM' do not mean the same in both cases. In the `odd' superlattices we observe passing from the configurations of type A-\Ab $\,$ to A-A, while in the `even' superlattices from AB-AB to AB-\Ab \Bb. Although it is hard to reason why we observe just these conigurations, once again we want to recall the cases of $\textrm{(MnTe)}_{2} \textrm{(ZnTe)}_{4}$ and $\textrm{(MnTe)}_{2} \textrm{(ZnTe)}_{6}$ systems, for which the results are different from those obtained for the respective superlattices with the thicker magnetic layers. They may suggest that not only surface effects play role in the process of the interlayer coupling and that contribution from the internal magnetic planes is so small that it manifests itself only in the case of weakly coupled systems.

In addition we have examined a series of tetragonally distorted \slat superlattices. They are characterized by the three parameters: the common in-plane lattice constant $\textrm{a}^{\perp}$ and the two lattice constants along the growth direction ($\textrm{a}_{\textrm{ZnTe}}^{\parallel}$, $\textrm{a}_{\textrm{MnTe}}^{\parallel}$). The quantities $\textrm{a}^{\perp}$ and $\textrm{a}_{\textrm{MnTe}}^{\parallel}$ have been taken from Ref.\ \onlinecite{GKSLFR1993}, while $\textrm{a}_{\textrm{ZnTe}}^{\parallel}$ has been calculated based on the known elastic constants $c_{11}$ and $c_{12}$ of ZnTe. Their relations to a lattice constant of the undistorted superlattice are the following:
\begin{itemize}
\item $\textrm{a}^{\perp} \approx 0.986 \textrm{a}_{\textrm{SL}}$
\item $\textrm{a}_{\textrm{ZnTe}}^{\parallel} \approx 0.971 \textrm{a}_{\textrm{SL}}$
\item $\textrm{a}_{\textrm{MnTe}}^{\parallel} \approx 1.056 \textrm{a}_{\textrm{SL}}$
\end{itemize}
According to Ref.\ \onlinecite{GKSLFR1993} strain plays an important role in determining magnetic structure of analysed superlattices. Specifically it governs the population of the domain types in the sample. Although we have taken into account only one-domain configurations we wanted to verify if changes in the lattice structure and its parameters could modify the calculated IEC essentially.
It turned out that these modifications have not introduced any qualitative changes to the IEC. Both nonmonotonic shape of the IEC and pace of its decrease have been preserved. As previously, the configurations with n=4,6 have displayed differences for m=2 and m>2. However, there have been quantitative changes. In all the cases, except for the superlattices with n=4, the strength of the IEC has increased or stayed unchanged for the distorted lattice. There has been no changes to the IEC for the systems with n=1,5, while for those with n=2,3,6,7 it has risen by factors ranging from 1.33 to 2.5. The only exception -- the superlattices with n=4 -- has shown approx. 91\% drop in the IEC and for m=2 it has changed the sign.
 
The last calculation concerned stability of the discussed magnetic configurations with respect to the rotation of the spins around the axis parallel to the growth direction. As we have already mentioned, helical structures were observed in other superlattices incorporating MnTe (e.g. MnTe/CdTe). Our target was to check if occurrence of the magnetic configuration of this type could lead to the lowering of the total energy. In fact we considered simpler situation, namely the system with all the spins rotated by the same angle $\phi$ in the second magnetic layer (which means that we took only the systems with the magnetic unit cell doubled with respect to the chemical one, i.e.\ those with n+m equal to an even number). The calculated total energy differences $\Delta E(\phi)$ between the system with rotated spins and the unperturbed one, for both the in-phase and antiphase configurations, revealed that the latter was a ground state in every examined case. As we were interested in looking for the energetic minimum only, we performed the calculation for the systems with the small values of the angle ($\phi \in (0^{\circ},10^{\circ})$). In all the cases $\Delta E(\phi)$ had the same -- parabolic -- shape with minimum at $\phi=0$. 
Undoubtedly these results cannot serve as a proof that the model we have used predicts that observed colinear structures are the ground states of MnTe/ZnTe superlattices. To prove this one would have to perform calculations for real helical structures and compare these results with the presented here. Unfortunately not all the structures can be calculated rigorously, as incommensurate phases require infinitely large unit cells. However, if it is true that the contribution to the IEC coming from internal magnetic layers is small (as our results suggest), then we can state that the above described structure with the whole layer rotated by common angle $\phi$ pretty well simulates helical structure.

\begin{acknowledgments}
We wish to gratefully acknowledge the guidance of Professor J. Blinowski throughout the course of the investigation. We also thank Dr P. Kacman and Professor W. Bardyszewski for critical reading of the manuscript.
\end{acknowledgments}

\end{document}